\documentstyle[11pt,newpasp,twoside,epsf]{article}
\def\edcomment#1{\iffalse\marginpar{\raggedright\sl#1\/}\else\relax\fi}

\reversemarginpar

\begin{document}
 \title{Dynamical processes, element mixing and
chemodynamical cycles in dwarf galaxies}
 \author{Andreas Rieschick and
Gerhard Hensler}
 \affil{Institut f\"ur Theoretische Physik und
Astrophysik, Universit\"at Kiel,
 D-24098 Kiel, Germany}
 	
\begin{abstract} 
Since the chemical evolution of galaxies seems to differ between morphological
types and deviates in many details from the standard scenario the question has 
to be addressed when, how and to what amount metal-enriched ejecta
from Supernovae and Planetary Nebulae polute their environment. Since
recent observations of dwarf galaxies show no significant metal abundance
gradients throughout the galaxies while enhancement of metals happens in
isolated HII regions, an effective mixing process has to be assumed. 
Chemodynamical evolution models can provide a possible explanation by
demonstrating that strong evaporation of gas clouds by hot gas and
following condensation leads to an almost perfect mixing of the gas. 
We focus on the different phases of chemodynamical evolution that are
experienced by a representative dwarf irregular galaxy model and present a 
quantitative analysis of the chemodynamical gas flow cycles.
\end{abstract}

\vspace{-5mm}

\section{Introduction} 
\vspace{-2mm}
Observations of dwarf galaxies show no significant metal abundance gradients
throughout the galaxies (Kobulnicky, Kennicutt, \& Pizagno 1999; van Zee et
al.\ 1998). A possible mechanism suggested by Tenorio-Tagle (1996) is the mixing
of metal-enriched blow-out gas with fresh matter from the environment of the
galaxy. After Supernovae (SNe) have caused an outflow of matter (depending on the
mass of the galaxy; see Mac Low \& Ferrara, 1998), the hot gas cools, condenses 
and falls back into the galactic body.

The moderate-to-low stellar metallicities in dwarf elliptical galaxies (dEs)
and the related dwarf spheroidals (dSphs) suggest that extensive gas loss has
occured during their evolution by means of SNe-driven galactic winds (Larson
1974; Dekel \& Silk 1986). Current starburst dwarf galaxies (SBDGs) are
characterized by superwinds (Marlowe et al.\ 1995) or by large expanding X-ray
plumes which are often confined by swept-up H$\alpha$ shells. Yet many dSph
galaxies show not only a significant intermediate-age stellar population
(Hodge 1989; Grebel 1997), but also more recent star formation (SF) events
(Smecker at al.\ 1994; Han et al.\ 1997) indicating that gas was partly kept
in the systems. On the other hand, gas infall might also cause a new SF episode
as in NGC 205 (Welch, Sage, \& Mitchell 1998). In several SBDGs large HI
reservoirs envelope the luminous galactic body (e.g. NGC 4449: Hunter et al.\
1998; I Zw 18: van Zee et al.\ 1998; NGC 1705: Meurer, Staveley-Smith, \&
Killeen 1998) and obtrude that the starburst is fueled by enhanced infall.
Since the infall rate cannot yet be evaluated observationally the task for
evolutionary models is to investigate the efficiency of SF and its according
gas consumption.\\[-5mm]

\section{The Model}
\vspace{-2mm}

\subsection{The chemodynamical description} 
Because of their low gravitational energy dwarf galaxies are strongly exposed
to the energetic impact from processes like stellar winds, supernovae or even
stellar radiation. When investigating gas mixing by large-scale dynamics and
by small-scale exchanges it is necessary to distinguish between the
dynamically separated gas components: the dense cloudy medium (CM) and the hot
dilute intercloud medium (ICM). To account for differences in energy and
timescale of the stellar yield an evolutionary code also has to divide stars
into different stellar mass components: high-mass stars (HMS), intermediate-mass
stars (IMS) and the not yet evolved low-mass stars (LMS).

The evolution calculation presented here was performed using our
chemodynamical evolution code CoDEx, a two dimensional grid-based
axisymmetrical evolution code including matter, momentum and energy
equations of all relevant processes in a self-consistant way. A detailed
description can by found in Samland, Hensler \& Theis (1997). Differing from
the specifications given there recent HMS yields by Woosley \& Weaver
(1995) are used which provide secondary nitrogen production (and a small
amount of primary nitrogen production). Also, for consistency, the most recent
IMS models by van der Hoek \& Groenewegen (1997) have been applied. 

Contrary to simple chemical evolutionary models that do not distinguish between
CM and ICM and therefore have to parameterize separation and mixing of
elements and possible selective outflow from the galaxy, the chemodynamical
description allows to trace the flow and mixing of metal enriched matter in a
self-consistent way.
\vspace{-2mm}

\subsection{Initial conditions}
We assume that the protogalactic gas cloud given by a Plummer-Kuzmin model
(Satoh 1980) has been formed in a cosmological CDM scenario but was prevented
from cooling by the metagalactic UV radiation field (Kepner, Babul, \&
Spergel, 1997). Not before this background radiation drops, the protogalactic 
gas cloud can cool due to recombination and subsequently collapes.

The total numerical grid size is $20\ kpc$ x $20\ kpc$, the spatial
resolution in the central parts amounts to almost $100\ pc$.
The evolved numerical model starts with a baryonic mass of about $10^{9}\
M_{\sun}$. A static dark matter halo according to 
Burkert (1995) with a mass of $10^{10} M_{\sun}$ is added.
This model is aimed to represent a dwarf irregular galaxy (dIrr).\\[-5mm]

\section{Results}
\vspace{-2mm}

\subsection{Dynamical phases of the evolution} 
For analytical purposes we devide the model galaxy into four equidistant zones
of radial extent of $0.5\ kpc$ along the galactic plane up to $2.0\ kpc$ and
with a z-height of $1.0\ kpc$. The four zones represent tori due to the
axisymmetrical grid. For consistency we use this scheme throughout the whole
evolution, even when the model galaxy changes its shape (e.g. by the settling 
of a disk-like structure in the galactic plane).

In Figure 1 the SF history for these four zones is plotted. The four curves
are stacked, i.e.\ the uppermost curve is the sum of all four zones. An
analysis of the model's kinematics (Rieschick \& Hensler, in prep.) implies five
distinct dynamical phases of evolution that are indicated in Fig.\ 1 by
vertical lines.

\begin{figure} 
\plotfiddle{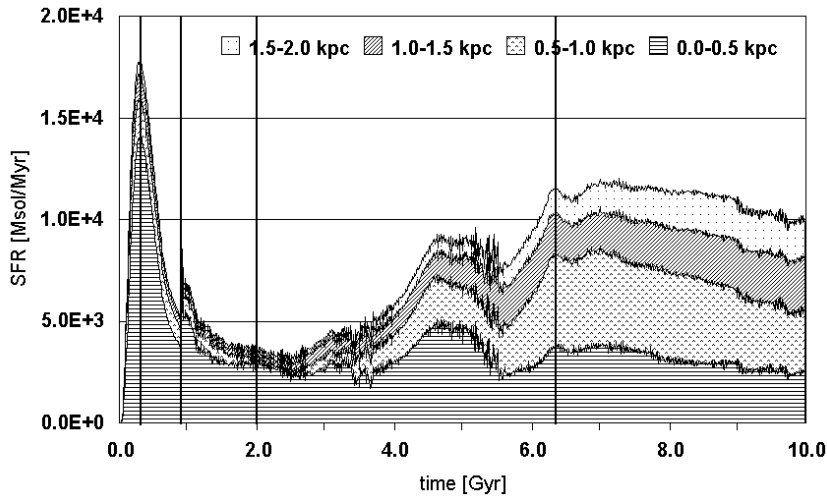}{6.00cm}{0}{37}{37}{-160}{0} 
\caption{\small Star formation
history of the $10^{9} M_{\sun}$ chemodynamical dIrr model in units of
$M_{\sun}/ Myr$ for different radial zones in the equatorial plane. The
absolute values correspond to the differences between two curves. The vertical
lines devide the different evolutionary phases described in the text.}
\end{figure}

{\it Collapse phase ($0$ - $0.3\ Gyr$):}\hspace{0.5cm} 
The protogalactic gas distribution is dynamically stable but cools and
collapses. The net gas infall rate amounts to $3.2 \cdot 10^{-1} M_{\sun} /
yr$ leading to a central density increase by a factor of 100. Thus the SF rate
rises steeply according to its quadratic dependence on the CM-density. The
succeeding supernova type II (SNII) explosions starting with a time delay of
about one $Myr$ smooth the further collapse of the CM. The peak in CM density
is reached after about $300\ Myr$ with a total SF rate of $1.8 \cdot 10^{-2}
M_{\sun} / yr$.

{\it Post-collapse phase ($0.3$ - $0.8\ Gyr$):}\hspace{0.5cm} 
Evaporation and gas outflow triggered by SNeII lead to further reduction of
CM density down to $1/4$ of the maximum value. The net gas loss rate averaged
over this phase amounts to $6.6 \cdot 10^{-2} M_{\sun} / yr$. A second much
smaller bounce occurs until an equilibrium between gravitation and gas
pressure is reached.

{\it Transitional phase ($0.8$ - $2.0\ Gyr$):}\hspace{0.5cm} 
In this phase the dynamical structure of the model galaxy changes completely.
While during the collapse and the post-collapse phase the inward or outward
motions, respectively, show almost coherence, now they differ locally and the
SF splits into several small regions. This process is accompanied by both
decreasing total gas density and total SF.

{\it Turbulent phase ($2.0$ - $6.2\ Gyr$):}\hspace{0.5cm} 
Two different zones have formed: the central part with a radius of about $0.5\ 
kpc$ and a thick disk in the range between $0.5$ and $2\ kpc$ distance from the
galactic center. These two parts follow separate paths in SF history. While
the central SF rate stays almost constant, the disk SF rate increases nearly by
a factor of ten in three steps. Thus the outer zones' SF rate reaches more
than $\twothirds$ of the total value while it has been only about $20 \%$ 
before. Regularly this zone splits into a few separate SF regions (Hensler \&
Rieschick 1999) while SF is wandering outwards in the disk to larger radial
distances because a growing number of gas clouds is accreted there due to 
infall.

{\it Irregular phase ($> 6.2\ Gyr$):}\hspace{0.5cm} 
A global quasi-stability seems to be reached now, even though the SF in the
outer parts of the galaxy happens in shortly living patches. Due to the
smaller gas density the exchange with the surrounding is larger than in the
central part leading to a non-existing abundance gradient. Because of the
incoherence and inhomogenities of dynamics and local regions we denote this
phase as "irregular". Figure 2 shows the SF rate at $8.0\ Gyr$ as an example.
While the highest SF rate per volume ($1.6 \cdot 10^{-6} M_{\sun}/
[Myr \cdot pc^{3}$])
exists in the center of the galaxy, three destinct SF regions with SF rates of
about $\onehalf$ of the previous value are visible in the disk. The SF decreases
slightly due to less infalling CM gas from the depleted HI reservoir.

\begin{figure} 
\plotfiddle{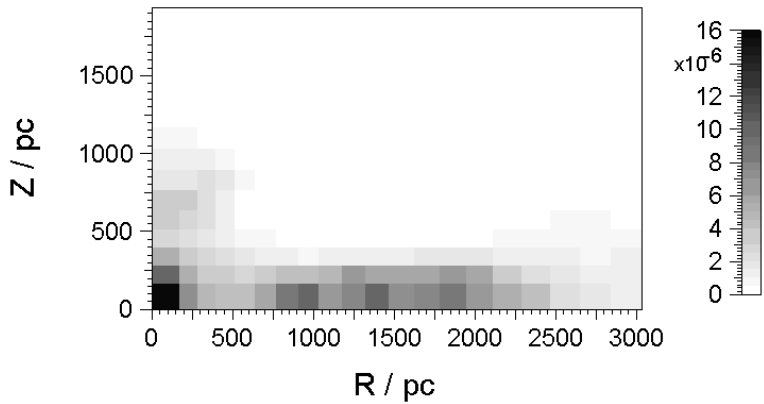}{5.50cm}{0}{40}{40}{-150}{0} 
\caption{\small 
Star formation rate at $8.0\ Gyr$ in the single numerical grid cells in units
of $M_{\sun}/(Myr \cdot pc^{3})$.
} \end{figure}
\vspace{-2mm}

\subsection{The gas flow cycle} 
Especially when studying the chemical evolution of galaxies the mixing and
exchange processes between gas phases become important. Matter is expelled by
SNe as hot dilute gas (ICM) and can return as clouds of cold gas (CM)
if it is gravitationally bound. The gas changes its state from ICM to CM by
condensation on the surface of gas clouds or vice versa by evaporation.

In Figure 3 we have analyzed the mass flow treated in chemodynamics between
the different components averaged between $6.2 - 10.0\ Gyr$ over a cylinder
with $r = 2\ kpc$ and $z = 1\ kpc$. From the mass flow rates one can
distinguish between an outer and an inner cycle. The outer one is produced by
infall of CM. From this about 24 \% is consumed by means of SF where 10 \% is
almost instantaneously rejected by HMS. The inner cycle represents SF and
stellar evolution and, therefore, contains different timescales.

The production of a minor fraction of hot ICM by SNe II leads to evaporation
of remaining CM and escapes from the galactic body as outflow as a consequence
of its high energy content. Significantly, almost 80 \% of the infall is
immediately converted into outflow by only 3.0 \% of gas that has gone through
the stellar cycle and is puffed up by stellar energy release. The full
evaporation rate can exceed the infall rate because shell sweep-up leads to
fragmentation and cloud formation in addition to a small amount of
condensation.

We emphasize that even though a large amount of the outflowing gas is
gravitationally unbound and leaves the galactic body, the metals produced in
HMS are for the most part kept in the outer gas flow cycle by mixing ICM with
continuously infalling clouds. As a result of this mechanism only a few
percent of the metals leave the gravitational field of the galaxy with the
outflowing ICM. While this mixing itself happens on a timescale of about $20\ Myr$,
the complete cycle of metal enrichment takes almost $1\ Gyr$ because of the
low infall velocity at later evolutionary stages. In contrary, the inner cycle
leads to an efficient self-enrichment of SF regions within $10\ Myr$ (see
also Skillman, Dohm-Palmer, \& Kobulnicky, 1998). The
outer enrichment timescale would be reduced significantly, however, in a 
scenario of a rapidly infalling intergalactic gas cloud.

\begin{figure} 
\plotfiddle{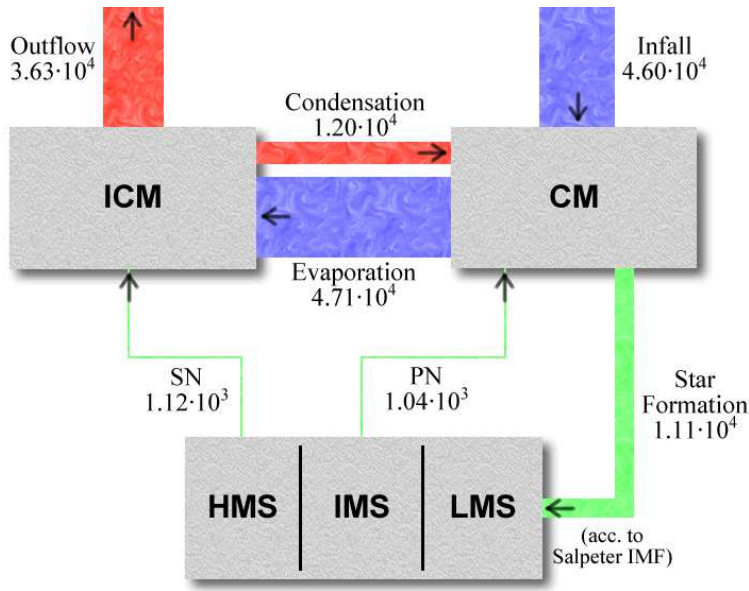}{7.5cm}{0}{40}{40}{-135}{0} 
\caption{\small Flow of matter
between the components in the chemodynamical dIrr model. The diagram shows the
temporal average over the interval $6.2$ to $10.0\ Gyr$ (the irregular phase, 
see text) for a radius of $2.0\ kpc$ and a z-height of $1.0\ kpc$, i.e.\ the whole
visible galaxy). All numbers are in units of $M_{\sun}/Myr$.} \end{figure}

This interaction scheme is running through all dynamical phases with different
amounts but nearly the same mass flux ratios. During the collapse phase
the infall rate is larger leading to intensive accumulation of matter in the
CM and stellar components. In the post-collapse phase the CM density is even
decreasing.\\[-5mm]

\section{Conclusions} 
\vspace{-2mm}
In the previous section we have shown that a constant infall of metal-poor
matter from the enveloping "HI reservoir" drives a permanent gas mixing cycle
keeping SNe-produced metals effectively in the gravitational field of the
galaxy. Since recent deep HI observations of dIrrs and SBDGs have discovered
an increasing number of large gas envelopes, i.e., of enormous gas reservoirs
circling or accumulating around the visible body of the BCDGs, the relevance
of infall episodes for the DG evolution is obvious and serves as the most
promissing explanation for their observed abundances (see Hensler, Rieschick,
\& K\"oppen, 1999).

Additionally we demonstrate by chemodynamical models that the SF cycle (Fig.
3) is triggered by only a small fraction of infalling matter, but produces
sufficient energy to cause major evaporation and to drive more than $3/4$ of
the infalling gas, incorporated into the ICM, as a galactic outflow back into
an outer long-term cycle. Chemodynamical models can provide a fundamental
insight into strong interactions between dynamical and energetical processes
that happen in these sensitively balanced systems of low gravitational
energy.\\[-1mm]

\begin{acknowledgements} {\small 
\setlength{\parskip}{-0.1ex}
We gratefully acknowledge cooperation and
discussions with J.~K\"oppen and Ch.~Theis. A.R. is supported by the {\it
Deutsche Forschungsgemeinschaft} under grant no. He 1487/23-1. The numerical
calculations have been performed at the computer center of the University of
Kiel and the NIC in J\"ulich. 
\setlength{\baselineskip}{3.0mm}
} \end{acknowledgements}\\[-5mm]


\begin{references} 
\vspace{-2mm}
{\small 
\setlength{\parskip}{-0.1ex}
\reference Burkert, A. 1995, \apj, 447, L25
\reference Dekel, A., \& Silk, J. 1986, \apj, 303, 39 
\reference Grebel, E.
  1997, Rev. Mod. Astron., 10, 29 
\reference Han, M., et al. 1997, \aj, 113, 1001
\reference Hensler, \& G., Rieschick, A. 1999, in Proc. XVIII Rencontre de  
Moriond, {\it Dwarf Galaxies and Cosmology}, eds. T. X. Thuan et al.  
(Gif-sur-Yvettes: Fronti\'eres), 461 
\reference Hensler, G., Rieschick, A., \&
  K\"oppen, J. 1999, in {\it The Evolution of Galaxies on Cosmological 
  Timescales}, eds. J. E. Beckman \& T. J. Mahoney, ASP Conference Series, 
  Vol. 187, 214
\reference Hodge, P. W. 1989, \araa, 27, 139 
\reference Hunter, D., et al.
  1998, \apj, 495, L47 
\reference Kepner, J. V., Babul, A., \& Spergel, D. N. 1997, \apj, 487, 61 
\reference Kobulnicky, H. A., Kennicutt, R. C., \&
  Pizagno, J. L. 1999, \apj,   514, 544 
\reference K\"oppen, \& J., Edmunds, M. G.
1999, \mnras, 306, 317 
\reference Larson, R. B. 1974, \mnras, 169, 229
\reference Mac Low, M. M., \& Ferrara, A. 1999, \apj, 513, 142
\reference Marlowe, A. T., Heckman, T. M., Wyse, R. F. G., \& Schomer, R. 1995,
  \apj, 438, 563 
\reference Meurer, G. R., Staveley-Smith, L., \& Killeen, N.
  E. B. 1998, \mnras, 300, 705 
\reference Satoh, C. 1980, \pasj, 32, 41
\reference Samland, M.,
  Hensler, G., \& Theis. Ch. 1997, \apj, 476, 544 
\reference Skillman, E. D., Dohm-Palmer, R. C., \& Kobulnicky, H. A. 1998, in
  {\it The Magellanic Clouds and Other Dwarf Galaxies}, proc. of the 
  Bonn/Bochum-Graduiertenkolleg Workshop, eds. T. Richler \& J. M. Braun, 
  Shaker Verlag, Aachen, p. 77
\reference Smecker-Hane, T. A.,
et al. 1994, \aj, 108, 507 
\reference Tenorio-Tagle, G. 1996, \aj, 111, 1641
\reference van den Hoek, L. B., \& Groenewegen, M.
  A. T. 1997, \aaps, 123, 305 
\reference van Zee, L., Westphal, D., Haynes, M.
  P., \& Salzer, J. J. 1998, \aj,   115, 1000 
\reference Welch, G. A., Sage, L. J.,
  \& Mitchell, G. F. 1998, \apj, 499, 209 
\reference Woosley, S. E., \& Weaver,T. A. 1995, \apjs, 101, 181
} 
\end{references}
\end{document}